\documentclass[runningheads]{llncs}
\usepackage[T1]{fontenc}
\usepackage{graphicx}
\begin{document}
\title{LIR: The First Workshop on \textbf{L}ate \textbf{I}nteraction and Multi Vector \textbf{R}etrieval @ ECIR 2026}
\titlerunning{LIR @ ECIR26}
%
\author{Benjamin Clavié\inst{1,7} \and
Xianming Li\inst{2,1} \and
Antoine Chaffin\inst{3} \and
Omar Khattab\inst{4} \and
Tom Aarsen\inst{5} \and
Manuel Faysse\inst{6} \and
Jing Li\inst{2}
}
\authorrunning{B. Clavié et al.}
%
\institute{{Mixedbread AI, San Francisco, California, USA \and
The Hong Kong Polytechnic University, Hong Kong SAR \and
LightOn, Paris, France \and
MIT, Cambridge, Massachussets, USA \and
HuggingFace, Netherlands \and
CentraleSupélec, Paris, France \and
National Institute of Informatics (NII), Tokyo, Japan
}\\
Correspondence: \email{ben@mixedbread.com}
}
\maketitle              
\begin{abstract}
Late interaction retrieval methods, pioneered by ColBERT, have emerged as a powerful alternative to single-vector neural IR. By leveraging fine-grained, token-level representations, they have been demonstrated to deliver strong generalisation and robustness, particularly in out-of-domain settings. They have recently been shown to be particularly well-suited for novel use cases, such as reasoning-based or cross-modality retrieval. At the same time, these models pose significant challenges of efficiency, usability, and integrations into fully fledged systems; as well as the natural difficulties encountered while researching novel application domains.
Recent years have seen rapid advances across many of these areas, but research efforts remain fragmented across communities and frequently exclude practitioners. The purpose of this workshop is to create an environment where all aspects of late interaction can be discussed, with a focus on early research explorations, real-world outcomes, and negative or puzzling results to be freely shared and discussed. The aim of LIR is to provide a highly-interactive environment for researchers from various backgrounds and practitioners to freely discuss their experience, fostering further collaboration.

\end{abstract}
\section{Motivation}

In recent years, like many application areas, Information Retrieval (IR) has experienced rapid developments due to the proliferation of deep learning-based systems, broadly referred to as ``Neural IR''~\cite{dpr,splade,monot5}. Neural network-based systems have achieved state-of-the-art effectiveness in many tasks~\cite{dpr,e5}. On the other hand, their rapid development has raised many research questions, with a large body of research focusing on better understand their working mechanisms, as well as efficiency concerns due to their computationally expensive nature clashing with the usually large corpora and low latency requirements of many IR tasks.

Among these developments, late-interaction multi-vector retrieval has emerged as a particularly promising area of research. Late interaction models, first introduced in the form of ColBERT~\cite{colbert} have demonstrated very strong performance, especially in out-of-domain settings, frequently outperforming other similarly-sized models. While the dominant approach to Neural IR involves representing both documents and queries as a single vector, late interaction operates at the token level, using one vector for each token and effectively representing both documents and queries as \textit{bags of tokens}. A document's relevance to a query is then computed via the maxsim operator, which compares every query token to every document token, keeping the highest score for each query token and finally summing them up to obtain a document-level score. This seemingly simple operator allows for fine-grained interaction between query and document terms, avoiding the information loss and associated performance decrease often observed when using single-vector methods. 

Outside academia, the industry adoption of late interaction methods has been staggered. While ColBERT became a common baseline in IR research and many papers explored it almost immediately after its introduction, widespread ``real-world'' adoption lagged behind, in large parts due to poor ecosystem integration compared to the simpler, more widespread single-vector methods. However, this rapidly changed in early 2023 with the rapid rise in the popularity of LLM Retrieval-Augmented Generation (RAG) pipelines combined with an increase in user-friendly tooling for ColBERT such as RAGatouille and PyLate, with ColBERT models soaring to millions of monthly downloads to the point of representing a noticeable proportion of HuggingFace download traffic.

In the same timeframe, research into late interaction models splintered into many streams. Some work showed that late interaction methods can seamlessly work in multimodal settings with the introduction of ColPali. Another large stream of research has focused on exploring methods to alleviate the significant efficiency constraints of late interaction, through various approaches: for example, more lightweight scoring objectives~\cite{xtr} and streamlined indexing methods~\cite{warp} and methods to storage footprint reduction including contributions from both academic~\cite{constbert,lightcolpali} and industrial~\cite{pooling,crisp}. Other areas of work, among many, have included exploring the applicability of ColBERT to lower-resource languages~\cite{jacolbertv2.5}, in multilingual~\cite{colbertxm} or cross-lingual~\cite{xcolbert} settings.

As this took place, industry practitioners began releasing an increasing number of new models, each time setting new state-of-the-art results and reaching large adoption~\cite{gtecolbert,colnomic,jina} while experimenting with novel training methods~\cite{jacolbertv2.5,colnomic} and exploring the impact of data mixes~\cite{jina}.

However, these streams of research, as well as many more such as studies dedicated to understanding the underlying mechanisms of ColBERT's query augmentation methods~\cite{masktokens}, have remained fractured while highly-related novel approaches such as Hyperencoders~\cite{hypencoder} have not had a space to be discussed from a late-interaction point of view. Much of the more applied work from industry has struggled to find suitable venues, particularly when their insight is small and focused while still being useful to the community at-large. 

At the moment, while Late Interaction methods are being increasingly used in both academic and industrial settings and shown to be particularly suitable for novel usecases centered on LLMs~\cite{reasongte} and as a way to bypass limitations inherent to single-vector representations~\cite{singlevec}, related publications are spread across a large number of publication venues, and there are limited opportunities for academic and industrial researchers to gather and exchange full-view insights on this rapidly developing area of research.

This workshop aims to serve as a bridge between all researchers currently working on various aspects of late interaction retrieval and multi-vector representations to learn from each other and engage in meaningful discussions bolstered by new perspectives. We aim for this workshop to foster greater integration for the community of researchers exploring this important area and provide new inspiration for future research.

\section{Theme and Purpose of the Workshop}

The LIR workshop has three main aims. The core goal \textbf{(1)} is to create a forum for researchers and practitioners from all communities interested in the exploration of late-interaction and multi-vector methods (``ColBERT'') to gather and freely discuss the challenges faced by current methods and the potential research avenues that would alleviate these challenges. Specifically, we hope for this space to serve as an important point of interaction between industry and academia. This is particularly important as robustness concerns have become a core area of interest in IR~\cite{robustir}, which requires the collaboration of practitioners from a diverse range of backgrounds to fully understand and address.  

Another major goal \textbf{(2)} is to serve as a space for researchers to discuss ongoing trends and early results, such as the growing importance of multi-modal retrievals and the necessity to adapt existing IR methods to novel uses of information retrieval such as Agentic Search and reasoning-model-powered retrieval. These are two areas where late-interaction models have shown promising early results~\cite{colpali,reasongte}, but also experience significant issues, such as poor handling of instructions~\cite{followir} and efficiency concerns, despite promising early efforts to alleviate them~\cite{lightcolpali}.  

Finally, \textbf{(3)} our hope is that this workshop will follow the long tradition of SIGIR and ECIR workshops inspiring future collaborative works bridging different research communities.  

\subsection{Topics of Interest}

Below is a non-exhaustive list of topics which have been identified as topics of interest during numerous discussions with researchers and practitioners who have expressed interest in a late-interaction workshop, and which will be specifically highlighted during our call for papers. For each area, we provide examples of past work which would fit within this category.  

\noindent \textbf{Late-Interaction Training Recipes} Studies on the impact of different scheduling, data mixes, training losses, and how they specifically impact late-interaction models, as previous work contains early signals that the impact of training decisions can vary significantly between single-vector and multi-vector retrievers~\cite{jacolbertv2.5}.  

\noindent \textbf{Theoretical Understanding of Late Interaction} Such as work that builds on the growing body of research focused on better understanding Chamfer similarity, which the MaxSim operator effectively calculates, and designing approximations with strong guarantees~\cite{muvera}, or building upon it to create more powerful approximators~\cite{LITE}.  

\noindent \textbf{Analysis of Late-Interaction-Specific Mechanisms} Late-interaction models rely on numerous tricks which have empirically been shown to improve performance but whose actual mechanisms are not fully understood. Previous work in this area has, for example, demonstrated that the use of [MASK] tokens for query augmentation did not function as ``generative augmentation,'' as previously thought~\cite{colbert}, but as ``term-weighters'' in a way similar to tf-idf weighting~\cite{masktokens}.  

\noindent \textbf{Multi-Modal Late Interaction} Multi-modal retrieval is an area becoming increasingly important due to the rapid development of multi-modal foundation models, both providing suitable backbones for the development of retrieval methods and enabling a large number of downstream applications. Early work in the area has suggested that late-interaction models are well suited to most modalities, with ColPali~\cite{colpali} and VideoColBERT~\cite{videcolbert} showing promising early results outperforming all other retriever families.  

\noindent \textbf{Alleviating Efficiency Concerns} The major drawback of late-interaction methods lies in their efficiency issues stemming from the need to store a considerable number of vectors. We specifically encourage contributions focused on highlighting the limits of current methods, which are frequently known to practitioners but poorly represented in the existing literature, as well as novel extensions of them to address these limits.  

\noindent \textbf{Usability} Due to their specificity, late-interaction models have often faced higher barriers to adoption due to not being compatible with the software and indexing stacks most practitioners are familiar with. General toolkits such as PyTerrier~\cite{terrier,pyterrier} and Anserini, as well as specialized libraries such as PyLate~\cite{pylate} and RAGatouille~\cite{ragatouille}, have greatly facilitated access, leading to greater adoption. However, there remains a large gap in fully facilitating access to the latest research methods, with novel scoring techniques such as XTR~\cite{xtr}, indexing methods like MUVERA~\cite{muvera}, or, more generally, multimodal approaches remaining hard to effectively implement.  

\noindent \textbf{Nascent Applications} With the rapid development of large language models, long-context retrieval capabilities, and better capabilities for Agentic Search—such as those measured by the new BrowseComp-Plus benchmark showing that retrievers have a large impact on so-called ``Deep Research'' tasks~\cite{browsecompplus}—and reasoning-based retrieval, where the model uses a reasoning LLM's thinking traces to perform fine-grained retrieval~\cite{reasonir}, have become topics of considerable interest. We welcome any contributions exploring these novel aspects and shining a light on, or improving, late-interaction methods' performance in these novel settings.

\section{Format and Structure}

\subsection{Workshop Format and Tentative Schedule}

For this first edition, we intend for the Late Interaction workshop to span half a day, focused largely on fostering open discussions rather than a series of non-interactive presentations, as shown in Table~\ref{tab:timetable}, which presents the tentative timetable. The workshop will be hybrid for all sessions except the roundtable discussion, to avoid communication issues caused by the hybrid aspect. The Keynote talk will cover the origin of late-interaction retrieval, its inception, and reflect on the development of the domain since.  

A short initial paper session will consist of oral presentations of recent pioneering work, while the second session will focus on presentations of demonstrations and ongoing work to feed into the Round Table discussion that follows, themed around future developments of multi-vector retrieval in the rapidly evolving landscape.  

Our intent with this format is to encourage interactions between various actors in the late-interaction ecosystem, from both academia and industry, and reflect on how learnings from different environments can be combined to inspire future research.  

\begin{table}[h!]
\centering
\caption{Workshop Timetable}
\label{tab:timetable}
\begin{tabular}{lc}
\hline
\textbf{Time} & \textbf{Activity} \\
\hline
\multicolumn{2}{l}{\textbf{Morning}} \\
\hline
09:00 -- 09:15 & Opening \\
09:15 -- 10:15 & Keynote -- Omar Khattab \\
10:15 -- 10:45 & Paper Session (Oral) \\
10:45 -- 11:00 & Coffee Break and breakout discussions \\
11:00 -- 11:45 & Paper Session (Demos and Posters) \\
11:45 -- 12:00 & Coffee Break and Round Table Preparation \\
12:00 -- 12:45 & Round Table Discussion: \textit{What next for Late Interaction?} \\
12:45 -- 13:00 & Reflections and Closing Notes \\
\hline
\end{tabular}
\end{table}

\section{Contributions}

\subsection{Types of Contributions}

We invite submissions in multiple forms: fully-fledged research papers, position papers, demo or technical reports, and opinion papers.  

We also welcome the submission of ongoing work and strongly encourage the sharing of both early and negative results to further the understanding of the underlying mechanisms of late-interaction methods. We will accept three submission formats:
\begin{itemize}
    \item \textbf{Notes} (up to 2 pages), which can be short, hyper-focused technical reports describing narrow results, positive or negative, or presenting ongoing work. Notes will all be presented as posters.
    \item \textbf{Papers} in both short (up to 4 pages) and full-length (up to 9 pages) formats, to be presented either orally or as a poster, as recommended by the program committee (PC).
    \item \textbf{Tech Reports} of a minimum of 2 pages and a maximum of 9 pages, describing publicly available software or models, to be presented as demos or posters.
\end{itemize}

All submissions will be reviewed by program committee members for relevance to the workshop and, more particularly, to the community at large. All papers may be submitted for archival, with proceedings published in CEUR-WS, or as non-archival. All notes will be archived by default, as their short page count ensures they will not constitute an obstacle to the publication of future longer versions.  

\section{Distinction from Main Conference Topics}

The core topic of this workshop, late-interaction retrieval, has overlap with the main topics of ECIR, as it is fundamentally an approach to information retrieval. However, late-interaction research currently sits at the intersection of many research communities, including the IR (ECIR, SIGIR), machine/deep learning (ICML, ICLR, NeurIPS), and NLP (EMNLP, ACL) communities. Due to this fragmentation, there currently is no forum to suitably discuss early results, even informally. We also seek to provide a strong space for practitioners to interact with researchers, by soliciting applied submissions that may not represent theoretical novelty but shed light on real-world constraints and usages. This workshop will complement ECIR by offering space for practical insights, early results, and industry experiences.  

\section{Organizers}

Our organizational committee is composed of a mix of industry and academic researchers with strong interests in multi-vector retrieval. Their previous work has explored the depth of this research area and highlighted many considerations, in terms of theory, evaluation methods, efficiency concerns, and potential impact of downstream applications. Their combined expertise provides a good overview of the many facets of multi-vector retrieval and will help spark interesting discussions.

\textbf{Benjamin Clavié} is a Researcher at Mixedbread Inc. and an incoming PhD student at the National Institute of Informatics (NII) in Tokyo. His research centers on encoder models as a whole, having initiated and co-led the ModernBERT project~\cite{modernbert}, with a specific focus on improving the usability of late interaction models. His work has helped popularize real-world use of ColBERT~\cite{ragatouille} and his exploration of training dynamics has become a standard way of training late-interaction models~\cite{jacolbertv2.5}, reaching considerable real-world adoption~\cite{hfdl}. He has served on the program committee for CIKM25 and workshops such as NLP4HR.

\textbf{Xianming Li} is a PhD candidate at Hong Kong Polytechnic University, under the supervision of Prof. Jing Li. He is also a researcher at Mixedbread Inc. His previous work has explored novel ways to improve the training of sentence representation models, highlighted by the development of methods such as the AnglE training loss~\cite{loss} and Espresso Sentence Embeddings~\cite{ese}, among others, with the resulting models enjoying large real-world adoption~\cite{hfdl}. His current interests focus on the use of late-interaction models for true multi-modal and multi-lingual retrieval. He has previously served on the program committee for a special session of BESC 2022. He is also a reviewer of ICLR/ACL/EMNLP/NAACL.

\textbf{Antoine Chaffin} is a researcher at LightOn. His research interests are in encoder models, having co-led the ModernBERT project~\cite{modernbert}, efficient information retrieval, and multimodal approaches. His recent work in the field of late-interaction retrieval has focused on increasing accessibility, creating PyLate~\cite{pylate}, a framework to support research into novel training methods, and contributing novel models such as GTE-ModernColBERT~\cite{gtecolbert} and methods such as Token Pooling~\cite{pooling}. Antoine has been awarded the French NLP Society (TAL)'s Best PhD Thesis Award. He has contributed to many large-scale research efforts, most notably BLOOM~\cite{bloom} and T0.

\textbf{Tom Aarsen} is a research engineer at Hugging Face. He is the lead developer and maintainer of the Sentence Transformers~\cite{st} library, the de facto library for practitioners to use sentence representation models of all kinds. He has been involved in a large number of collaborations with both academic and industry partners to facilitate the dissemination of their work and ensure they can be faithfully reproduced. Tom is a key representative of industry needs. 

\textbf{Manuel Faysse} is a PhD student at CentraleSupélec. His research has focused on training and industrial applications of open large language models. He introduced the ColPali~\cite{colpali} family of late-interaction models, demonstrating the suitability of pre-trained Vision-Language Models as backbones for natively multimodal late-interaction models. His subsequent work has explored other applications of late-interaction retrieval, in contexts such as contextualized retrieval~\cite{inteb}. 

\textbf{Jing Li} is an Associate Professor at Hong Kong Polytechnic University (PolyU). She established and currently leads the PolyU Embodied Artificial Intelligence Lab under PolyU-COMP. She has broad research interests in NLP, human-centered AI, and Embodied AI. She has been an active member of the NLP community and has served in senior positions as an organization committee member and area chair for major conferences such as ACL, NAACL, and EMNLP.

\textbf{Omar Khattab} is an Assistant Professor at the Massachusetts Institute of Technology (MIT). His research interests have broached many topics, with a specific focus on information retrieval and research into the creation of end-to-end AI systems centered on Large Language Models via the DSPy project~\cite{dspy}. Omar's research has pioneered modern multi-vector retrieval, with the original ColBERT approach~\cite{colbert} as well as ColBERTv2~\cite{colbertv2} and the de facto standard PLAID indexing method~\cite{plaid}. Omar has received numerous awards for his research work, most recently a SIGIR 2025 Best Paper Award; he has given numerous invited talks and has served as a Senior Area Chair for ARR and a PC member for many venues, among which NeurIPS, ICLR, ICML, and TOIS.

\section{Audience}

\textbf{Expected Audience} We expect the workshop to have around 30 attendees. The target audience is composed of researchers from both industry and academia, as well as industry practitioners who typically have fewer opportunities to engage with academic communities. We intend to provide a forum for researchers from various communities (ML/DL, NLP, IR) with an interest in multi-vector methods to discuss future research avenues.

\textbf{Dissemination} We will promote the workshop on social media platforms, most notably X, and conduct a strong outreach campaign by directly reaching out to individual researchers and engineers, as well as to companies interacting with the field of late interaction. We will set up a dedicated website for the workshop and actively seek submissions. We are already in contact with many practitioners who have expressed interest in attending this workshop should it take place.

%
%
%
\bibliographystyle{splncs04}
\bibliography{bib}

\begin{thebibliography}{10}
\providecommand{\url}[1]{\texttt{#1}}
\providecommand{\urlprefix}{URL }
\providecommand{\doi}[1]{https://doi.org/#1}

\bibitem{bloom}
BigScienceWorkshop, Scao, T.L., Fan, A., Akiki, C., Pavlick, E., Ili{\'c}, S., Hesslow, D., Castagn{\'e}, R., Luccioni, A.S., Yvon, F., et~al.: Bloom: A 176b-parameter open-access multilingual language model. arXiv preprint arXiv:2211.05100  (2022)

\bibitem{gtecolbert}
Chaffin, A.: Gte-moderncolbert (2025), \url{https://huggingface.co/lightonai/GTE-ModernColBERT-v1}

\bibitem{reasongte}
Chaffin, A.: Reason-moderncolbert (2025), \url{https://huggingface.co/lightonai/Reason-ModernColBERT}

\bibitem{pylate}
Chaffin, A., Sourty, R.: Pylate: Flexible training and retrieval for late interaction models. arXiv preprint arXiv:2508.03555 (Accepted at CIKM 2025, to be published)  (2025)

\bibitem{browsecompplus}
Chen, Z., Ma, X., Zhuang, S., Nie, P., Zou, K., Liu, A., Green, J., Patel, K., Meng, R., Su, M., et~al.: Browsecomp-plus: A more fair and transparent evaluation benchmark of deep-research agent. arXiv preprint arXiv:2508.06600  (2025)

\bibitem{jacolbertv2.5}
Clavi{\'e}, B.: Jacolbertv2. 5: Optimising multi-vector retrievers to create state-of-the-art japanese retrievers with constrained resources. Journal of Natural Language Processing  \textbf{32}(1),  176--218 (2025)

\bibitem{ragatouille}
Clavi{\'e}, B.: Ragatouille: Easily use and train state-of-the-art late-interaction retrieval methods (colbert) in any rag pipeline. \url{https://github.com/AnswerDotAI/RAGatouille} (2025), version 0.0.9 (2025-02-10)

\bibitem{pooling}
Clavi{\'e}, B., Chaffin, A., Adams, G.: Reducing the footprint of multi-vector retrieval with minimal performance impact via token pooling. arXiv preprint arXiv:2409.14683  (2024)

\bibitem{inteb}
Conti, M., Faysse, M., Viaud, G., Bosselut, A., Hudelot, C., Colombo, P.: Context is gold to find the gold passage: Evaluating and training contextual document embeddings. arXiv preprint arXiv:2505.24782  (2025)

\bibitem{colpali}
Faysse, M., Sibille, H., Wu, T., Omrani, B., Viaud, G., Hudelot, C., Colombo, P.: Colpali: Efficient document retrieval with vision language models. In: ICLR (2025)

\bibitem{splade}
Formal, T., Piwowarski, B., Clinchant, S.: Splade: Sparse lexical and expansion model for first stage ranking. In: Proceedings of the 44th International ACM SIGIR Conference on Research and Development in Information Retrieval. pp. 2288--2292 (2021)

\bibitem{masktokens}
Giacalone, B., Paiement, G., Tucker, Q., Zanibbi, R.: Beneath the [mask]: An analysis of structural query tokens in colbert. In: European Conference on Information Retrieval. pp. 431--439. Springer (2024)

\bibitem{LITE}
Jain, H., Ji, Z., Rawat, A.S., Veit, A., Jayasumana, S., Reddi, S.J., Menon, A.K., Yu, F.: Efficient document ranking with learnable late interactions. In: 2nd Workshop on Advancing Neural Network Training: Computational Efficiency, Scalability, and Resource Optimization (WANT@ ICML 2024) (2024)

\bibitem{muvera}
Jayaram, R., Dhulipala, L., Hadian, M., Lee, J.D., Mirrokni, V.: Muvera: Multi-vector retrieval via fixed dimensional encoding. Advances in Neural Information Processing Systems  \textbf{37},  101042--101073 (2024)

\bibitem{dpr}
Karpukhin, V., Oguz, B., Min, S., Lewis, P.S., Wu, L., Edunov, S., Chen, D., Yih, W.t.: Dense passage retrieval for open-domain question answering. In: EMNLP (1). pp. 6769--6781 (2020)

\bibitem{dspy}
Khattab, O., Singhvi, A., Maheshwari, P., Zhang, Z., Santhanam, K., Haq, S., Sharma, A., Joshi, T.T., Moazam, H., Miller, H., et~al.: Dspy: Compiling declarative language model calls into state-of-the-art pipelines. In: The Twelfth International Conference on Learning Representations (2024)

\bibitem{colbert}
Khattab, O., Zaharia, M.: Colbert: Efficient and effective passage search via contextualized late interaction over bert. In: Proceedings of the 43rd International ACM SIGIR conference on research and development in Information Retrieval. pp. 39--48 (2020)

\bibitem{hypencoder}
Killingback, J., Zeng, H., Zamani, H.: Hypencoder: Hypernetworks for information retrieval. In: Proceedings of the 48th International ACM SIGIR Conference on Research and Development in Information Retrieval. pp. 2372--2383 (2025)

\bibitem{xtr}
Lee, J., Dai, Z., Duddu, S.M.K., Lei, T., Naim, I., Chang, M.W., Zhao, V.: Rethinking the role of token retrieval in multi-vector retrieval. Advances in Neural Information Processing Systems  \textbf{36},  15384--15405 (2023)

\bibitem{loss}
Li, X., Li, J.: Aoe: Angle-optimized embeddings for semantic textual similarity. In: Proceedings of the 62nd Annual Meeting of the Association for Computational Linguistics (Volume 1: Long Papers). pp. 1825--1839 (2024)

\bibitem{ese}
Li, X., Li, Z., Li, J., Xie, H., Li, Q.: Ese: Espresso sentence embeddings. In: The Thirteenth International Conference on Learning Representations, ICLR2025 (2025)

\bibitem{robustir}
Liu, Y.A., Nachimovsky, H., Zhang, R., Kurland, O., Guo, J., Tennenholtz, M.: Robust-ir@ sigir 2025: The first workshop on robust information retrieval. In: Proceedings of the 48th International ACM SIGIR Conference on Research and Development in Information Retrieval. pp. 4142--4145 (2025)

\bibitem{colbertxm}
Louis, A., Saxena, V.K., van Dijck, G., Spanakis, G.: Colbert-xm: A modular multi-vector representation model for zero-shot multilingual information retrieval. In: Proceedings of the 31st International Conference on Computational Linguistics. pp. 4370--4383 (2025)

\bibitem{lightcolpali}
Ma, Y., Li, J., Zang, Y., Wu, X., Dong, X., Zhang, P., Cao, Y., Duan, H., Wang, J., Cao, Y., et~al.: Towards storage-efficient visual document retrieval: An empirical study on reducing patch-level embeddings. arXiv preprint arXiv:2506.04997  (2025)

\bibitem{constbert}
MacAvaney, S., Mallia, A., Tonellotto, N.: Efficient constant-space multi-vector retrieval. In: European Conference on Information Retrieval. pp. 237--245. Springer (2025)

\bibitem{pyterrier}
Macdonald, C., Tonellotto, N., MacAvaney, S., Ounis, I.: Pyterrier: Declarative experimentation in python from bm25 to dense retrieval. In: Proceedings of the 30th acm international conference on information \& knowledge management. pp. 4526--4533 (2021)

\bibitem{xcolbert}
Nair, S., Yang, E., Lawrie, D., Duh, K., McNamee, P., Murray, K., Mayfield, J., Oard, D.W.: Transfer learning approaches for building cross-language dense retrieval models. In: Proceedings of the 44th European Conference on Information Retrieval (ECIR) (2022), \url{https://arxiv.org/abs/2201.08471}

\bibitem{monot5}
Nogueira, R., Jiang, Z., Pradeep, R., Lin, J.: Document ranking with a pretrained sequence-to-sequence model. In: Findings of the Association for Computational Linguistics: EMNLP 2020. pp. 708--718 (2020)

\bibitem{terrier}
Ounis, I., Amati, G., Plachouras, V., He, B., Macdonald, C., Johnson, D.: Terrier information retrieval platform. In: European Conference on Information Retrieval. pp. 517--519. Springer (2005)

\bibitem{videcolbert}
Reddy, A., Martin, A., Yang, E., Yates, A., Sanders, K., Murray, K., Kriz, R., de~Melo, C.M., Van~Durme, B., Chellappa, R.: Video-colbert: Contextualized late interaction for text-to-video retrieval. In: Proceedings of the Computer Vision and Pattern Recognition Conference. pp. 19691--19701 (2025)

\bibitem{st}
Reimers, N., Gurevych, I.: Sentence-bert: Sentence embeddings using siamese bert-networks. arXiv preprint arXiv:1908.10084  (2019)

\bibitem{plaid}
Santhanam, K., Khattab, O., Potts, C., Zaharia, M.: Plaid: an efficient engine for late interaction retrieval. In: Proceedings of the 31st ACM International Conference on Information \& Knowledge Management. pp. 1747--1756 (2022)

\bibitem{colbertv2}
Santhanam, K., Khattab, O., Saad-Falcon, J., Potts, C., Zaharia, M.: Colbertv2: Effective and efficient retrieval via lightweight late interaction. In: Proceedings of the 2022 Conference of the North American Chapter of the Association for Computational Linguistics: Human Language Technologies. pp. 3715--3734 (2022)

\bibitem{warp}
Scheerer, J.L., Zaharia, M., Potts, C., Alonso, G., Khattab, O.: Warp: An efficient engine for multi-vector retrieval. In: Proceedings of the 48th International ACM SIGIR Conference on Research and Development in Information Retrieval. pp. 2504--2512 (2025)

\bibitem{reasonir}
Shao, R., Qiao, R., Kishore, V., Muennighoff, N., Lin, X.V., Rus, D., Low, B.K.H., Min, S., Yih, W.t., Koh, P.W., et~al.: Reasonir: Training retrievers for reasoning tasks. arXiv preprint arXiv:2504.20595  (2025)

\bibitem{colnomic}
Team, N.: Nomic embed multimodal: Interleaved text, image, and screenshots for visual document retrieval (2025), \url{https://nomic.ai/blog/posts/nomic-embed-multimodal}

\bibitem{crisp}
Veneroso, J., Jayaram, R., Rao, J., {\'A}brego, G.H., Hadian, M., Cer, D.: Crisp: Clustering multi-vector representations for denoising and pruning. arXiv preprint arXiv:2505.11471  (2025)

\bibitem{e5}
Wang, L., Yang, N., Huang, X., Jiao, B., Yang, L., Jiang, D., Majumder, R., Wei, F.: Text embeddings by weakly-supervised contrastive pre-training. arXiv preprint arXiv:2212.03533  (2022)

\bibitem{modernbert}
Warner, B., Chaffin, A., Clavi{\'e}, B., Weller, O., Hallstr{\"o}m, O., Taghadouini, S., Gallagher, A., Biswas, R., Ladhak, F., Aarsen, T., Adams, G.T., Howard, J., Poli, I.: Smarter, better, faster, longer: A modern bidirectional encoder for fast, memory efficient, and long context finetuning and inference. In: Che, W., Nabende, J., Shutova, E., Pilehvar, M.T. (eds.) Proceedings of the 63rd Annual Meeting of the Association for Computational Linguistics (ACL 2025) (Volume 1: Long Papers). pp. 2526--2547. Association for Computational Linguistics, Vienna, Austria (Jul 2025)

\bibitem{singlevec}
Weller, O., Boratko, M., Naim, I., Lee, J.: On the theoretical limitations of embedding-based retrieval (2025), \url{https://arxiv.org/abs/2508.21038}

\bibitem{followir}
Weller, O., Chang, B., MacAvaney, S., Lo, K., Cohan, A., Van~Durme, B., Lawrie, D., Soldaini, L.: Followir: Evaluating and teaching information retrieval models to follow instructions. In: Proceedings of the 2025 Conference of the Nations of the Americas Chapter of the Association for Computational Linguistics: Human Language Technologies (Volume 1: Long Papers). pp. 11926--11942 (2025)

\bibitem{hfdl}
Wolf, T.: Most liked and most downloaded open-source ai models from 2022 to 2024. \url{https://huggingface.co/posts/thomwolf/250854638539377} (Dec 2024), hugging Face post, posted 4 December 2024

\bibitem{jina}
Xiao, H., Wang, B., Jha, R.: Jina-colbert-v2: A general-purpose multilingual late interaction retriever. In: Proceedings of the Fourth Workshop on Multilingual Representation Learning (MRL 2024). pp. 159--166 (2024)

\end{thebibliography}
%




\end{document}